\documentstyle[12pt,fleqn]{article}
\begin{document}
\title{DIELECTRIC PROPERTIES OF THE QUASI-TWO-DIMENSIONAL ELECTRON
 LIQUID IN HETEROJUNCTIONS}
\author{{C. BULUTAY}\\ Department of Electrical and Electronics
 Engineering, \\ Middle East Technical University, Ankara - 06531,
 TURKEY\\ and \\  M. TOMAK \\ Department of Physics, \\ Middle
 East Technical University, Ankara - 06531, TURKEY.}
\date{} 
\maketitle
\renewcommand\baselinestretch{2}
\begin{abstract}
A quasi-two-dimensional (Q2D) electron liquid (EL) is formed at the
 interface of a semiconductor heterojunction. For an accurate
 characterization of the Q2D EL, many-body effects need to be taken
 into account beyond the random phase approximation. In this
 theoretical work, the self-consistent static local-field correction
 known as STLS is applied for the analysis of the Q2D EL. The
 penetration of the charge distribution to the barrier-acting
 material is taken into consideration through a variational
 approach. The Coulomb from factor that describes the effective 2D
 interaction is rigorously treated. The longitudinal dielectric
 function and the plasmon dispersion of the Q2D EL are
 presented for a wide range of electron and ionized acceptor
 densities choosing GaAs/AlGaAs as the
 physical system. Analytical expressions fitted to our results are
 also supplied to enable a widespread use of these results.
\end{abstract}
{\bf PACS:} 71.45.Gm, 73.20.Mf
\renewcommand\baselinestretch{2}
\small\normalsize
\section{INTRODUCTION}
The name {\it electron liquid} or {\it electron gas} refers to a
 model system formed by interacting dynamical electrons within a
 medium containing a uniformly distributed positive charge having
 no motion. The overall system is electrically neutral. As the
 positive background is rigid, it does not respond to any kind of
 excitation, hence, it cannot polarize, however, the electrons can.
 The
 three-dimensional (3D) electron liquid (EL) has been studied as a
 model system for metals \cite{sing-tosi} and the 3D positive ion
 liquid was proposed as a model astrophysical system \cite{star}. 
 In the case of two-dimensions, the study of the two-dimensional
 (2D) EL has been driven mainly by technological advances such as
 silicon inversion layers \cite{ando}, modulation doped field
 effect transistors \cite{hiyam}, intercalated graphite layers
 \cite{graphite}, and fractional quantum Hall effect in 2D electron
 systems \cite{fqhe}. In addition to its technological importance
 the 2D EL contains rich physics due to enhanced particle
 correlations and geometrical parameters that characterize the
 actual realization of the 2D system.

The EL stayed as a problem of interest in the past few decades and
 intense research efforts lead to several advances in the field. For
 the 2D case the first major contribution was due to Stern who
 calculated the density-density response function of the
 noninteracting EL \cite{stern67}, which is known in the 3D case as
 the Lindhard function. The Stern function (i.e., 2D Lindhard
 function) immediately made the random phase approximation (RPA)
 available to 2D EL. RPA was at that time one of the most successful
 many-body approaches for the EL. In 1976, Jonson \cite{jonson}
 showed that for 2D EL, a many-body approach proposed by Singwi and
 coworkers \cite{singwi1} (referred to as STLS) performed remarkably
 better than RPA. We have very recently compared the 2D-STLS
 technique with the quantum Monte Carlo data of Tanatar and Ceperley
 \cite{tanatar89} and proposed analytical forms for the dielectric
 function of the {\it ideal} 2D EL based on the STLS technique
 \cite{bulutay96}.

The knowledge of the dielectric function and the local-field
 correction paves the way for a variety of many-body related terms
 such as self-energy, carrier lifetime, mobility \cite{mahan1}.
 Connections to  density-functional theory can also be established
 \cite{DFT}. Furthermore, the dielectric screening plays a
 substantial role in the characterization of other excitations,
 such as, polarons \cite{dsarma83}, \cite{dsarma85}, \cite{devr86}.

In this work, our aim is to present an accurate and systematic
 characterization of the dielectric properties of the quasi-2D (Q2D)
 EL in real heterojunctions where the electron distribution can
 penetrate to both sides of the interface. The charge distribution
 is based on a variational approach proposed by Bastard
 \cite{bastard}. The effective 2D electron interaction for this
 system is characterized by the Coulomb form factor. This quantity
 is treated rigorously. The dielectric function for the GaAs/AlGaAs
 heterojunction is given for wide ranges of electron and ionized
 acceptor densities. Throughout this work the term {\it dielectric
 function}, refers to the {\it longitudinal} dielectric function.
 We also fitted analytical expressions to our
 data for the efficient use of these results by other researchers.
 To simplify the computational labour we stayed in the
 zero-temperature formulation and the so-called electrical quantum
 limit, where only the lowest subband along the confinement
 direction is populated (we refer to a very recent work
 \cite{velicia}, discussing the effects of higher subbands on
 the dielectric function).

The paper is planned as follows: Sec.~II discusses briefly the
 variational computation of the Q2D electron distribution. The
 effective 2D interaction of these Q2D electrons is treated in
 Sec.~III and the modifications to the STLS technique in the Q2D
 case is contained in Sec.~IV. In Secs.~V and VI the dielectric
 function and the plasmon dispersion are considered respectively;
 all the results are given referring to GaAs/AlGaAs  as the
 physical system, however, the approach is developed for a general
 heterojunction. In Sec.~VII, the fitted analytical expressions for
 the results are presented. Following the conclusion section,
 appendices include some details on variational formulation and the
 Coulomb form factor for a Q2D system.

\section{VARIATIONAL CHARGE DISTRIBUTION FOR A HETEROJUNCTION}
The electrons from ionized donors in the barrier side of a
 modulation doped heterojunction are trapped in a wedge-like well
 formed by a step barrier due to conduction band edge discontinuity
 on one side, and the potential due to presence of the transferred
 electrons and ionized acceptors on the other \cite{ando}. The
 one-dimensional quantum confinement gives the Q2D nature to the
 system and behaves remarkably different than ideal 2D and 3D
 systems. In handling the many-body effects in heterojunctions we
 avoid some critical simplifications that have been used in the
 past such as infinite barrier height \cite{dsarma83},
 \cite{dsarma85}, \cite{devr86} (which is a reasonable
 approximation only for Si inversion layers) and no ionized
 acceptors within the channel \cite{gold} (which is in fact not the
 case in practise). For an accurate account of the electronic
 distribution in heterojunctions, we  use Bastard's variational
 approach that was tested previously in determining the subband
 energies \cite{bastard}.

The electronic wave function $\varsigma_{i}(z)$, within the
 effective mass approximation satisfies the one-dimensional
 Schr\"{o}dinger equation along the confinement direction (chosen to
 be the $z$-direction)
\begin{equation}
\left [ \frac{-\hbar^{2}}{2} \frac{d}{dz} \frac{1}{m(z)} \frac{d}
{dz} +U_{e-e}(z)+U_{A}(z)+U_{barrier}(z) \right ] \varsigma_{i}
(z)=E_{i}\, \varsigma_{i}(z) .
\end{equation}
$U_{e-e}(z)$ is the potential (energy) formed by the presence of the
 electrons, $U_{A}(z)$ is the potential due to ionized acceptors and
 $U_{barrier}(z)$ is a step-barrier potential: $U_{barrier}(z)=U_b\,
 \Theta (-z)$, resulting from the conduction band edge mismatch of
 the neighbouring materials. $m(z)$ is the effective mass of the
 conduction band electrons being equal to $m_B$ in the
 barrier-acting material and $m_A$ in the well-acting material.
 Bastard proposed the following variational form for the lowest
 subband $\varsigma_{1}(z)$ allowing penetration to the barrier
 region, ($z<0$) \cite{bastard}
\begin{equation}
\label{eq:var_wf}
\mbox{\large $\varsigma_1$}(z)=\left \{
\begin{array}{ll}
M e^{\kappa_{b}z/2},&\mbox{for $z\leq 0$}  \\
N (z+z_{0}) e^{-bz/2},&\mbox{for $z\geq 0$}
\end{array} \right. .
\end{equation}
 Invoking the continuity of $\varsigma_{1}(z)$ and \begin{math}
 m^{-1}(z) \frac{d}{dz} \varsigma_{1}(z)\end{math}
 at $z=0$ and the normalization of $\varsigma_{1}(z)$, \begin{math}
 \int_{-\infty}^{+\infty} dz|\varsigma_{1}(z)|^2 \, =1  \end{math},
 yields the following three equations,
\begin{equation}
\label{eq:M=}
M=N z_{0} ,
\end{equation}
\begin{equation}
z_{0}=\frac{2}{b+\kappa_{b} \frac{m_{A}}{m_{B}}} ,
\end{equation}
\begin{equation}
\label{eq:N=}
N=\sqrt{\frac{b^{3}}{2 \left [1+bz_{0}+\frac{b^{2}z_{0}^{2}}{2}
 \left (1+\frac {b}{\kappa_{b}} \right ) \right ] }} .
\end{equation}

Bastard also set $\kappa_b=2 \sqrt{2m_{B}U_{b}/\hbar^{2}}$ and used
 $b$ as the only variational parameter. We have observed that such a
 choice of $\kappa_b$ is highly satisfactory in the electrical
 quantum limit. Note that $M, N$ and $z_0$
 also depend on $b$ through Eqs.(\ref{eq:M=})-(\ref{eq:N=}). $b$ is
 determined by minimizing the {\it total} system energy (see
 Appendix~A for the expressions). A closed form representation of
 $b$ is not possible, unlike the Si-inversion layer \cite{ando},
 however, the minimization can easily be achieved numerically.  We
 work in the regime where only the lowest subband is populated, this
 puts an upper limit to the 2D electron density above which the
 Fermi level crosses the first-excited subband energy. For
 GaAs/AlGaAs heterojunction our analysis is valid for the 2D
 electronic densities, $N_{2D}\leq 7\times 10^{11}\,
 \mbox{cm}^{-2}$. Bastard's work \cite{bastard} can be consulted
 for further details.

\section{COULOMB FORM FACTOR FOR A PENETRABLE HETEROJUNCTION}
In the 2D EL the interaction potential in reciprocal space is taken
 to be $2 \pi e/q$, where $q$ is the wave number. This potential is
 obtained by taking the 2D Fourier transform of the 3D Coulomb
 interaction which is $1/R$, $R$ denoting distance in real space
 \cite{iwamoto}. In fact a strictly 2D solution of Poisson's
 equation is proportional to $-\ln (R)$ \cite{barton} rather than
 $1/R$ and its 2D Fourier transform is proportional to $1/q^2$ as
 in 3D EL. However, the $-\ln (R)$ interaction is seldom used
 \cite{lnr} due to indication by real physical 2D systems that
 $1/R$ type of interaction is relevant \cite{olego},\cite{isih}.
 For the case of a Q2D system the charge distribution along the
 third dimension modifies the effective 2D interaction from
 $2 \pi e/q$ to $F(q) \, 2 \pi e/q$. $F(q)$ is the Coulomb form
 factor describing the effect
 of the finite spread of the charge distribution along the
 confinement direction over a region where the background
 dielectric constant is discontinous due to different materials on
 both sides.

Following the approach in the previous section we use the
 variational charge distribution that can leak into the barrier
 region and calculate the function $F(q)$ accordingly. The details
 on $F(q)$ are given in the Appendix~B, here we state the final
 result,
\begin{equation}
\label{eq:Fq}
F(q)=\frac{1}{2} \left ( 1+\frac{\epsilon_{B}}{\epsilon_{A}}
 \right ) I_{1}+\frac{1}{2} \left ( 1-\frac{\epsilon_{B}}
{\epsilon_{A}} \right ) I_{2}+\frac{1}{2} \left ( 1+
\frac{\epsilon_{A}}{\epsilon_{B}} \right ) I_{3}+\frac{1}{2}
 \left ( 1-\frac{\epsilon_{A}}{\epsilon_{B}} \right ) I_{4}+2I_{5} ,
\end{equation}
where,
\begin{eqnarray}
\label{eq:I1}
I_{1}=2 N^{4} \left \{ \frac{1}{2b} \left [ \frac{z_{0}^{4}}{(b+q)}+
\frac{2z_{0}^{3}}{(b+q)^{2}}+\frac{2z_{0}^{2}}{(b+q)^{3}} \right ]
+\frac{1}{(2b)^{2}} \left [ \frac{4z_{0}^{3}}{(b+q)}+
\frac{6z_{0}^{2}}{(b+q)^{2}}+\frac{4z_{0}}{(b+q)^{3}} \right ]
 \right .\nonumber\\  \left .
+\frac{1}{(2b)^{3}} \left [ \frac{12z_{0}^{2}}{(b+q)}+
\frac{12z_{0}}{(b+q)^{2}}+\frac{4}{(b+q)^{3}} \right ]+
\frac{1}{(2b)^{4}} \left [ \frac{24z_{0}}{(b+q)}+
\frac{12}{(b+q)^{2}} \right ]+\frac{1}{(2b)^{5}} \frac{24}{(b+q)}
 \right \} ,
\end{eqnarray}
\begin{equation}
I_{2}=N^{4} \left [ \frac{z_{0}^{2}}{(b+q)}+\frac{2z_{0}}
{(b+q)^{2}}+\frac{2}{(b+q)^{3}} \right ]^{2} ,
\end{equation}
\begin{equation}
I_{3}=\frac{M^{4}}{\kappa_{b} (\kappa_{b}+q)} ,
\end{equation}
\begin{equation}
I_{4}=\frac{M^{4}}{(\kappa_{b}+q)^{2}} ,
\end{equation}
\begin{equation}
\label{eq:I5}
I_{5}=\frac{M^{2} N^{2}}{(\kappa_{b}+q)} \left [ \frac{z_{0}^{2}}
{(b+q)}+\frac{2z_{0}}{(b+q)^{2}}+\frac{2}{(b+q)^{3}} \right ] .
\end{equation}
In Eq.~(\ref{eq:Fq}) $\epsilon_A$ and $\epsilon_B$ are the
 background dielectric constants of the well-acting and the
 barrier-acting materials respectively. The bare electron-electron
 interation potential energy for this Q2D system becomes
\begin{equation}
U^{Q2D}(q)=\frac{2 \pi e^{2}}{\bar \epsilon\, q} F(q)
\end{equation}
where $\bar \epsilon =(\epsilon_{A}+\epsilon_{B})/2$ and $q$ is the
 2D wave number associated with the spatial variation along the 2D
 sheet.

The terms containing $I_2$ and $I_4$ in Eq.~(\ref{eq:Fq}) represent
 the image interaction resulting from the different permittivities
 on both sides. Their effects decrease when the permittivity
 contrast diminishes; an example is the GaAs/AlGaAs system
 considered in Fig.~1 for two different electron densities (see the
 following section for the material parameters used). The Coulomb
 form factor becomes more important in high electron densities (see
 Fig.~1) where the in-plane particle separation is comparable
 to the extension of the charge distribution along the confinement
 direction. The expression for $F(q)$ in Eq.~(\ref{eq:Fq}) will
 especially be useful for heterojunctions with a high permittivity
 difference and a low barrier height.
 
\section{Q2D STLS}
The STLS technique in 2D has been discussed in the literature, and
 we refer, for instance, to Jonson's pioneering paper \cite{jonson}.
 In going from 2D to Q2D the only modification (within the
 electrical quantum limit) is the replacement of
 the 2D Coulomb interaction energy by the effective 2D interaction
 due to finite extension of the charge distribution along the
 confinement direction. The exchange and correlation hole
 associated with each electron in the system is described by the
 local-field correction, $G(q)$. This function in the case of Q2D
 STLS reads
\begin{equation}
\label{eq:Gofq1}
G^{Q2D}(q)=\int\!\!\int \frac{d^2 p_n}{2 \pi} \,\frac{F(p)}{F(q)}\,
\frac{\vec{q_n} \cdot\vec{p_n}}{q_n \, p_n}\, \left[1-S(|\vec{p}
 - \vec{q}|) \right] ,
\end{equation}
where the subscript $n$ is used in this equation and in the rest of
 the text to denote wave numbers normalized to the Fermi wave number
 $k_F$ (i.e., $q_n\equiv q/k_F$ etc.). In Eq.~(\ref{eq:Gofq1}) $F$
 is the Coulomb form factor and $S$ is the static structure factor.
 The latter contains contributions from plasmons and electron-hole
 pairs and is related to the dielectric function through the
 fluctuation-dissipation theorem. The dielectric function, in turn,
 depends on the local-field correction (see Sec. V). The
 computational task involves the self-consistent solution of these
 three coupled nonlinear integral equations. A change of variables
 leads to a substantial improvement in the execution speed of the
 STLS algorithm. Using $\vec{t_n}=\vec{p_n}-\vec{q_n}$ in Eq.~
(\ref{eq:Gofq1}) leads to
\begin{equation}
\label{eq:Gofq2}
G^{Q2D}(q)=\frac{1}{\pi \, F(q)} \int_{0}^{\infty} dt_n\, t_n\,
 [1-S(t)] \int_{0}^{\pi} d\phi \, F\left ( q \sqrt{1+a^2 +2a\cos
 \phi} \right )\frac{a\cos \phi +1}{\sqrt{1+a^2 +2a\cos \phi}} ,
\end{equation}
where $a=t_n/q_n$.

In Fig.~2 we present the self-consistent STLS $G^{Q2D}(q)$ results
 for a wide range of electronic densities given in terms of $r_s$.
 $r_s$ is the effective interparticle spacing defined as $r_{s}=
1/a_{B}^{*} \sqrt{\pi N_{2D}}$ where $N_{2D}$ is the 2D electronic
 density and  $a_{B}^{*}$ is the effective Bohr radius given by
 \begin{math} a_{B}^{*}=\frac{\bar \epsilon}{m^{*}} \frac{\hbar^{2}}
{m_{0} e^{2}} \end{math}, $\bar \epsilon$ is the background average
 static dielectric constant and $m^{*} m_0$ is the effective mass of
 the electrons considered, with $m_0$ being the free electron mass.
  We consider GaAs/AlGaAs heterojunction as the physical system with
 the parameters $m_{A}=0.07 m_{0}$, $m_{B}=0.088 m_{0}$,
 $\epsilon_{A}=13$, $\epsilon_{B}=12.1$ and $U_{b}=0.3\, \mbox{eV}$
 (corresponding to an Al mole fraction of 0.3) which were used by
 Stern and Das Sarma \cite{stern84}. For $a_{B}^{*}$ we used
 $\bar \epsilon=12.55$ and $m^{*}=0.07$, giving $a_{B}^{*}=9.49\,
 \mbox{nm}$. The  conduction band offset, $U_{b}$ was measured by
 some groups to be around $0.225\, \mbox{eV}$ (in contrast to $0.3\,
 \mbox{eV}$) \cite{barrier}. We have observed that our results are
 not sensitive to the deviation of $U_{b}$ in this range. In Fig.~2
 the interval $r_{s}=0.8-20$ is shown with an ionized acceptor
 density of $N_{depl}=0.46\times 10^{11}\, \mbox{cm}^{-2}$. For
 $r_{s}<0.8$ the higher subbands start to be populated which was not
 taken into our analysis.

 For the 2D EL, STLS $G(q)$ becomes proportional to $q$ as
 $q\rightarrow 0$ \cite{bulutay96} whereas in 3D case it is
 proportional to $q^2$ \cite{sing-tosi}. In the Q2D case we
 observe that (see Fig.~2) for low $r_s$ values small-$q$ behavior
 is close to quadratic and as $r_s$ increases this behavior goes
 towards a linear one indicating an approach to a 2D character.

Gold and Calmels also reported their results on $G^{Q2D}(q)$ for
 GaAs/AlGaAs heterostructure \cite{gold}. Their treatment is based
 on STLS but with essential discrepancies compared to ours. They
 imposed the local-field correction for 2D and Q2D to be of the form
\begin{equation}
\label{eq:G_GC}
 G_{GC}^{Q2D}(x)=r_{s}^{2/3} \frac{1.402\, x}{[\/2.644\,C_{12}^{2}
(r_{s})+x^{2} C_{22}^{2}(r_{s})]^{1/2}}
\end{equation}
 where \begin{math} x=\frac{q}{k_{F}}\frac{1}{\sqrt{2} r_{s}^{1/3}}
 \end{math} and the coefficients $C_{12}$ and $C_{22}$ were
 tabulated \cite{gold}. They assumed no penetration to barrier
 region in the Coulomb form factor and also neglected the presence
 of ionized acceptors in the well-acting region. Especially, the
 form used in Eq.~(\ref{eq:G_GC}) enabled them to reduce the
 computational effort appreciably, however, their results are in
 strong disagreement with ours for $r_{s}\geq 1$ and $q \simeq
 2k_F$ both in 2D \cite{bulutay96} and  Q2D as can be seen in
 Fig.~3. The form in Eq.~(\ref{eq:G_GC}) cannot accommodate the
 full STLS $G(q)$  leading to a poor dielectric function and
 screening properties. The ionized acceptors in the well region play
 a primary role and need to be included in the treatment.

 \section{DIELECTRIC FUNCTION}
The function of practical importance is the wave number- and
 frequency-dependent (longitudinal) dielectric function,
 $\epsilon(q,\omega)$ that not only determines the response to
 a weak external perturbation
 but also possesses information on the many-body dynamics of the
 system. With the knowledge of the local-field correction,
 $\epsilon(q,\omega)$ is given as
\begin{equation}
\label{eq:diel}
\mbox{\Large $\epsilon$}^{Q2D}_{STLS}(q,\omega)=\frac{1-U^{Q2D}(q)\,
 \mbox{\Large $\pi^0$}(q,\omega)\, [1-G^{Q2D}(q)]}{1+U^{Q2D}(q)\,
 \mbox{\Large $\pi^0$}(q,\omega)\, G^{Q2D}(q)} ,
\end{equation}
 where $\mbox{\Large $\pi ^0$}(q, \omega)$ is the 2D zeroth-order
 polarization insertion, the Stern function \cite{stern67},
 \cite{bulutay96}. Apart from $\mbox{\Large $\pi^0$}$, 2D and
 Q2D quantities behave differently. This is
 illustrated in Fig.~4 showing inverse static dielectric function,
 $\epsilon^{-1}(q,0)$ within RPA and STLS for both 2D and Q2D cases.
 To assess the effect of penetration of the charge distribution into
 the barrier region, we compare $U_{b}=0.1\, \mbox{eV}$ case with
 $U_{b} \rightarrow \infty$ in Fig.~5 at $r_{s}=0.8$. It is observed
 that for GaAs/AlGaAs-like heterojunctions, this penetration has a
 minor effect on the static dielectric function. In Fig.~6 the
 inverse static dielectric function of GaAs/AlGaAs heterojunction is
 plotted in the density range $r_{s}=0.8-20$ and for $N_{depl}=
0.46\times 10^{11}\, \mbox{cm}^{-2}$. Notably, the GaAs/AlGaAs
 heterostructure shows an overscreening effect (i.e., $\epsilon
 <0$) for $r_s \geq 3$. The onset of overscreening shifts to higher
 electron densities for the strictly 2D case \cite{bulutay96}, due
 to enhanced particle correlations in lower dimensions. As an
 interesting consequence, the negative dielectric function suggests
 a negative compressibility of the Q2D EL \cite{mahan1} and in fact,
 recently this has been experimentally observed on a GaAs quantum
 well structure \cite{eisenstein}.

We would like to include some necessary remarks about this
 dielectric function.
 The expression in Eq.~(\ref{eq:diel}) only gives the Q2D EL
 dielectric function. The total screened electron-electron
 interaction is
\begin{equation}
U^{Q2D}_{scr}(q, \omega)= F(q) \frac{2 \pi e^{2}}{\bar \epsilon\, q}
 \frac{1}{\epsilon ^{Q2D}_{STLS}(q,\omega)} .
\end{equation}
The dielectric responses of the polar lattice and the valence
 electrons are contained in the average background dielectric
 constant $\bar \epsilon$. Here we have used the {\it static}
 dielectric constant (see, for instance, our definition of $a_B^*$ in
 Sec.~IV), hence, it is assumed that the polar lattice can follow
 the external excitations. Obviously this limits the validity range
 of this work to $\omega \ll \omega_{TO}$, with $\omega_{TO}$ being
 the transverse optical phonon frequency. This limitation is relaxed
 if the background lattice does not have a polar character. Hence,
 for the particular system that we are considering , the dielectric
 function is expected to be valid up to about 1 THz. In principle,
 however, the static nature of the local field correction of the
 STLS technique can further limit this upper frequency.

Finally, the dielectric function given by Eq.~(\ref{eq:diel}) takes
 into account the polarization of the electrons  in the lowest
 subband. Even though the presently available experiments on
 GaAs/AlGaAs systems mainly fall into this regime \cite{heit89},
\cite{baum96}, the technological trend aims to populate the higher
 subbands to increase the amount of current carried in modulation
 doped field effect transistors by using different materials such
 as InGaAs/InAlAs \cite{hiyam}. When the higher subbands are
 occupied the dielectric function should necessarily be a tensor
 of the form $\epsilon_{ij}(q,\omega)$,
 where $i=j$ terms account for the intrasubband polarizations and
 $i \not= j$ terms represent intersubband couplings. To assess the
 performance of the presented approach regarding the electrical
 quantum limit, we extended the variational wave function technique
 to include lowest two subbands and determined the subband
 populations by invoking self-consistency between Poisson and
 Schr\"{o}dinger equations. In Fig.~7 we show the charge
 distributions along the confinement direction for a density of
 $1 \times 10^{12}\, \mbox{cm}^{-2}$. The solid curve represents the
 correct charge distribution containing contributions from the lowest
 and first-excited subbands. The dashed curve, on the other hand,
 sticks to the electrical quantum limit which actually breaks down
 beyond $N_{2D}= 7 \times 10^{11}\, \mbox{cm}^{-2}$. It is
 important to observe that the difference between the two curves is
 quite marginal. This is simply because the percentage of the
 first-excited subband electrons is 4.7\% at this density.

\section{PLASMON DISPERSION}
The elementary excitations in electron liquids are electron-hole
 pair creations and collective excitations knowns as plasmons
 \cite{pines}. The latter can be characterized with the
 knowledge of the wave number
 and frequency-dependent dielectric function, $\epsilon
 (q,\omega)$. Particularly, the plasmon dispersion relation,
 $\omega_p(q)$ is available through the zeros of the dielectric
 function;
\begin{equation}
\mbox{\Large $\epsilon$}(q,\omega_p(q))=0 .
\end{equation}
 Inserting the expression for $\epsilon (q,\omega)$ from
 Eq.~(\ref{eq:diel}) leads to the following closed form expression
 for the plasmon dispersion
\begin{equation}
\label{eq:plasmon}
\mbox{\Large $\nu_p$}(q)=\frac{q_n (z+1)}{2} \sqrt{q_n^2 +
 \frac{4}{z^2+2z}} ,
\end{equation}
where
\begin{equation}
z=\frac{q_n}{\sqrt{2} r_s\, F(q)\, [1-G^{Q2D}(q)]} ,
\end{equation}
and
\begin{equation}
\mbox{\Large $\nu_p(q)$}=\frac{\hbar \omega_p(q)}{2 E_F}=\frac{m
 \omega_p(q)}{\hbar k_F^2} .
\end{equation}
which is valid in the range $[0,q_{n,max}]$ where $q_{n,max}$
 satisfies \begin{math} \nu_p(q_{n,max}) = q_{n,max} +
 q_{n,max}^2/2 \end{math} and outside this region plasmons
 dissociate to electron-hole pairs so that collective excitations
 are no longer long-lived. The Eq.~(\ref{eq:plasmon}) reduces to the
 ideal 2D result \cite{mahan85} when $F(q) \rightarrow 1$. Fig.~8
 shows the plasmon dispersion for GaAs/AlGaAs heterostructure with
 $N_{depl}=0.46\times 10^{11} \mbox{cm}^{-2}$ and for several $r_s$
 values. Even though the plasmon dispersion can be experimentally
 probed, such as, through far infrared spectroscopy \cite{heit89},
 the available experimental results pertain to high electronic
 densities and small wave numbers $(q<k_F)$. Therefore the effects
 of the local-field correction have not yet been verified.

\section{ANALYTICAL EXPRESSIONS}
In this section we present our fitted expressions to $G^{Q2D}(q)$
 and $F(q)$ applicable to GaAs/AlGaAs heterojunction in the density
 range $r_{s}=0.8-20$. As a fit to $G^{Q2D}(q)$ (shown in Fig.~2 by
 solid lines), we tried a simple form containing three fitting
 parameters,
\begin{equation}
\label{eq:Gfit}
G_{fit}^{Q2D}(q)=A\left(1-e^{-\frac{B}{A} q_n^C} \right),
\end{equation} 
where $A,B$ and $C$ are the fitting parameters. The optimized values
 are tabulated in Table~I for $N_{depl}=0.46\times 10^{11}\,
 \mbox{cm}^{-2}$. The third parameter, $C$ is introduced based on
 our observations on the long-wavelength behavior of $G^{Q2D}(q)$ in
 Sec. IV. In ideal 2D, $C$ was equal to one and in 3D case $C$ was
 equal to two. Optimized $C$ values in Table~I show this
 interpolation between $r_{s}=0.8$ to 5, but then this trend is lost
 to enable a good fit for the whole $q$ values. The fitted
 expressions are plotted in Fig.~2 by the dotted lines. To assess
 the quality of the fitting we use the following error estimate
 between a target vector, $T(i)$ and the fitted vector $T_{fit}(i)$:
\begin{equation}
\label{eq:error}
error(\%)=\frac{1}{N}\sum_{i=1}^{N} \left | \frac{T(i)-T_{fit}(i)}
{T(i)}\right | \, 100 \, .
\end{equation}
Accordingly the deviation of the fitting in Fig.~2 is less than 2.5
 \%.

The Coulomb form factor, $F(q)$ also requires a laborious work for
 GaAs/AlGaAs system. This function can be fitted by a simple
 expression

\begin{equation}
\label{eq:Ffit}
F_{fit}(q)=\frac{1}{1+D q_n},
\end{equation}
containing a single fitting parameter $D$ which is tabulated in
 Table~I for the same $N_{depl}$ value.

 The knowledge of $G^{Q2D}_{fit}(q)$ and $F_{fit}(q)$ is sufficient
 for representing the dielectric function (see Eq.~(\ref{eq:diel})).
 The performance of fitting for $\epsilon^{-1}(q,0)$ is available
 from Fig.~6 (shown by dotted lines) where the error, using the
 estimate in Eq.~(\ref{eq:error}) is less than  1\%. Similarly in
 Fig.~8 the plasmon dispersions with the use of the fitted forms are
 shown in dashed lines, the fitting error being much less than
 0.1\%.

 We have observed that taking the barrier height $U_{b}=0.225\,
 \mbox{eV}$ does not significantly affect the parameters $A,B,C,$
 and $D$. However, $N_{depl}$ takes an important part in both $G(q)$
 and $F(q)$, so we repeated the self-consistent Q2D STLS technique
 for $N_{depl}=0.146,\, 1.47,\, 4.69\times 10^{11}\, \mbox{cm}^{-2}$
 and performed again fittings. Rather than specifying these results
 in tabular form we present below fitted {\it functions\/} of
 $r_{s}$ for $A,B,C$, and $D$.
\begin{equation}
\label{eq:Afit}
A_{fit}=1.02\left[1-a_{1}r_{s}^{a_{2}} e^{-a_{3}r_{s}} \right] ,
\end{equation}
\begin{equation}
B_{fit}=b_{1} \ln (b_{2} r_{s})+b_{3} ,
\end{equation}
\begin{equation}
C_{fit}=0.42 r_{s}^{-c_{1}}+1.03\, r_{s}^{0.12} ,
\end{equation}
\begin{equation}
\label{eq:Dfit}
D_{fit}=\frac{d_{1}}{d_{2}+r_{s}^{1.15}} ,
\end{equation}
the constant parameters contained in these expressions are tabulated
 in Table~II for the considered range of $N_{depl}$ values. With the
 expressions in Eqs.~(\ref{eq:Afit})-(\ref{eq:Dfit}), inverse static
 dielectric function can be generated to an accuracy of about 1 \%
 except $r_s=2$ case having an error about 9 \%. Similarly with
 these equations plasmon dispersion can be recovered to an error
 much less than 0.1 \%.

\section{CONCLUSION}
The dielectric properties of the Q2D EL in a heterostructure are
 studied and the behaviour is seen to be remarkably different than
 the strictly 2D EL \cite{bulutay96}. The analysis is rigorous with
 the only simplifications being the electrical quantum limit and the
 zero-temperature formalism. These simplifications can also be
 relaxed at the expense of computational complexity. The leakage of
 the charge distribution to the barrier region is included in the
 analysis through a variational approach. The full form of the
 Coulomb form factor applicable to a general heterostructure is 
 presented. For the GaAs/AlGaAs system, the image terms have been
 observed to have a marginal role. A sizeable contribution will be
 encountered in the case of heterostructures built up of materials
 having a high dielectric constant contrast and a low conduction
 band offset. The dielectric function and the plasmon dispersion of
 the Q2D EL are characterized using the STLS many-body approach that
 leads to substantial improvement over the conventional RPA.
 Unfortunately, in contrast to the ideal 2D case \cite{tanatar89},
 \cite{moroni92}, quantum Monte Carlo simulations are not available,
 to compare our results, for the Q2D EL; the present experimental
 data cannot cover the regime where the RPA breaks down (i.e.,
 $r_s>1$ and $q\simeq k_F$). Our analysis extends to a wide range
 of electron and ionized acceptor densities. To the best of our
 knowledge this work forms the most elaborate study of the screening
 properties of the Q2D EL. Our results are supplemented with
 analytical expressions fitted to our data. We have presented the
 expressions for the local-field correction, Coulomb form factor,
 the dielectric function and the plasmon dispersion of the Q2D EL.
 With this information, the self-energy, quasiparticle lifetime,
 mobility etc., can also be obtained; polarons in Q2D systems can be
 studied with the inclusion of electron-electron screening.

\section*{ACKNOWLEDGMENTS}
We gratefully acknowledge discussions with N. G\"{u}nalp, A.
 G\"{o}kalp, and K. Leblebicio\u{g}lu.
\pagebreak
\section*{APPENDIX~A: TOTAL SYSTEM ENERGY IN THE VARIATIONAL
 APPROACH}
In this section, for completeness we include the expression for the
 {\it total\/} system energy of a heterojunction in Bastard's
 variational approach \cite{bastard}. The ground-state expectation
 of the total system energy per electron is
\begin{equation}
\label{eq:Etot}
\langle\mbox{\~E}_{TOT}(b)\rangle=\langle T(b)\rangle +\frac{1}{2}
 \langle U_{e-e}(b)\rangle+\langle U_A(b)\rangle+\langle
 U_{barrier}(b)\rangle,
\end{equation}
where \begin{math}\langle T(b)\rangle\end{math} is the kinetic
 energy term given by
\begin{equation}
\langle T(b)\rangle=-\frac{M^2 \kappa_b}{4\, m_B^*} +\frac{N^2}
{2\, m_A^*\, b}\, ( 1+ bz_0-b^2 z_0^2/2) \;\mbox{Ry},
\end{equation}
\begin{math}\langle U_{e-e}(b)\rangle\end{math} is the average
 electron-electron interaction potential,
\begin{eqnarray}
\langle U_{e-e}(b)\rangle & = & \frac{8\pi}{\bar{\epsilon}}
 N_{2D} \,\left [ \frac{N^4}{b^7} \left ( \frac{33}{4}+
\frac{25 bz_0}{2}+\frac{17 b^2 z_0^2}{2}+3b^3z_0^3+
\frac{b^4z_0^4}{2} \right ) \right. \nonumber\\ & & \left.
 \mbox{} - \frac{N^2 M^2}{\kappa_b^2 b^3} (z_0^2b^2+2bz_0+2)
 \right ] \,\mbox{Ry},
\end{eqnarray}
where \begin{math}\langle  U_A(b)\rangle\end{math} is the average
 electron-ionized acceptor interaction potential,
\begin{equation}
\langle  U_A(b)\rangle=\frac{8\pi}{\bar{\epsilon}} N_{depl} \,
\left [ \frac{6 N^2}{b^4} \left ( 1+\frac{2}{3} bz_0 +
\frac{b^2 z_0^2}{6} \right ) -\frac{M^2}{\kappa_b^2} \right ]
 \;\mbox{Ry},
\end{equation}
where \begin{math}\langle U_{barrier}(b)\rangle\end{math} is the
 average potential energy due to step-barrier,
\begin{equation}
\langle U_{barrier}(b)\rangle=\frac{U_{b,Ry} N^2 z_0^2}{\kappa_b}
 \;\mbox{Ry},
\end{equation}
where \begin{math}\kappa_b=2\sqrt{m_B^* U_{b,Ry}}\end{math}. In
 above equations atomic units are used; all energies are in Rydbergs
 $(1 Ry=m_0 e^4/2\hbar^2)$ and lengths in Bohr radius
 $(a_B=\hbar^2/m_0 e^2)$. In the {\it total} system energy, the
 electron-electron interaction is weighted by 1/2 to avoid
 double-counting. The variational parameter $b$ is determined by
 minimizing $\langle\mbox{\~E}_{TOT}(b)\rangle$ which is an easy
 task numerically as the cost function has a single minimum.
\pagebreak
\section*{APPENDIX~B: EFFECTIVE 2D COULOMB INTERACTION}
We first recall the electrostatic potential due to a point charge
 $Q$, at a distance $d$ (along the $z$-axis) from the interface
 formed by two semi-infinite dielectric media with permittivities
 $\epsilon_{A}$ and $\epsilon_{B}$. Solution of the Poisson's
 equation subject to continuity requirements at the interface, $z=0$
 results in an electrostatic potential of the form \cite{jackson},
\begin{equation}
\mbox{\Large $\Phi$}(\vec R ,z)=\left\{
\begin{array}{ll}
\frac{1}{\epsilon_{A}} \left ( \frac{Q}{\sqrt{R^2 +(d-z)^2 }}+
\frac{\epsilon_{A}-\epsilon_{B}}{\epsilon_{A}+\epsilon_{B}} 
\frac{Q}{\sqrt{R^2 +(d+z)^2 }} \right ),& z \geq 0 \\ \\
\frac{2}{\epsilon_{A}+\epsilon_{B}}
\frac{Q}{\sqrt{R^2 +(d-z)^2 }},& z \leq 0 
\end{array} \right. .
\end{equation}
This result will now be used in constructing the effective 2D
 Coulomb interaction energy between two (charge) distributions
 $n(\vec R,z)$ and $n(\vec R',z')$
\begin{eqnarray}
\label{eq:app_U1}
\lefteqn{U^{Q2D}(\vec R-\vec R')= } \nonumber\\ & &
\frac{e^2}{\epsilon_A} \left\{
\int_0^\infty dz \int_0^\infty dz' \left[ \frac{n(\vec R,z)\:
 n(\vec R',z')}{\sqrt{|\vec R-\vec R'|^2+(z-z')^2}}
+\frac{\epsilon_A -
\epsilon_B}{\epsilon_A +\epsilon_B} 
\frac{n(\vec R,z)\: n(\vec R',z')}{\sqrt{|\vec R-\vec R'|^2
 +(z+z')^2}} \right ] \right \} \nonumber\\ & & 
 +\frac{e^2}{\epsilon_B} \left\{ \int_{-\infty}^0 dz \int_{-\infty}
^0 dz' \left[ \frac{n(\vec R,z)\: n(\vec R',z')}
{\sqrt{|\vec R-\vec R'|^2+(z-z')^2}} +\frac{\epsilon_B -
\epsilon_A}{\epsilon_A +\epsilon_B} \frac{n(\vec R,z)\:
 n(\vec R',z')}{\sqrt{|\vec R-\vec R'|^2+(z+z')^2}} \right ]
 \right \} \nonumber\\ & &
 +\frac{2e^2}{\epsilon_A +\epsilon_B} \left\{ \int_0^\infty dz
 \int_{-\infty}^0 dz' \frac{n(\vec R,z) \: n(\vec R',z')}
 {\sqrt{|\vec R-\vec R'|^2+(z-z')^2}} +\int_{-\infty}^0 dz
 \int_0^\infty dz' \frac{n(\vec R,z)\: n(\vec R',z')}
 {\sqrt{|\vec R-\vec R'|^2+(z-z')^2}} \right \} \nonumber\\ & &
\end{eqnarray}
The first two terms in Eq.~(\ref{eq:app_U1}) represent direct and
 image interaction of the charge distributions on the right side of
 the interface $(z>0)$. Third and fourth terms represent the same
 interactions for $z<0$ region. The last two terms which are in fact
 equal, represent the direct interaction between charge
 distributions on opposite sides of the interface.

The charged-particle distribution is $n(\vec R,z)\equiv n(z)=
|\varsigma_1 (z)|^2$, where $\varsigma_1 (z)$ is given in
 Eq.~(\ref{eq:var_wf}). The 2D Fourier transform of $U^{Q2D}
(\vec R-\vec R')$ is easily obtained using the result
\begin{equation}
\int d^2 r \frac{e^{-i\vec q \cdot \vec r}}{\sqrt{r^2 +a^2}}=
\frac{2\pi}{q} e^{-|a|q}
\end{equation}
as
\begin{equation}
\label{eq:app_U2}
U^{Q2D}(q)=\frac{2\pi e^2}{q \epsilon_A} \left(I_1 +
 \frac{\epsilon_A -\epsilon_B}{\epsilon_A +\epsilon_B}I_2 \right) +
\frac{2\pi e^2}{q \epsilon_B} \left(I_3 + \frac{\epsilon_B -
\epsilon_A}{\epsilon_A +\epsilon_B}I_4 \right) +\frac{8 \pi e^2}{q
 (\epsilon_A +\epsilon_B)} I_5,
\end{equation}
with,
\begin{equation}
I_1=\int_0^\infty dz \int_0^\infty dz' N^4 (z+z_0)^2 (z'+z_0)^2
 e^{-bz} e^{-bz'} e^{-|z-z'|q},
\end{equation}
\begin{equation}
I_2=\int_0^\infty dz \int_0^\infty dz' N^4 (z+z_0)^2 (z'+z_0)^2
 e^{-bz} e^{-bz'} e^{-(z+z')\/q},
\end{equation}
\begin{equation}
I_3=\int_{-\infty}^0 dz \int_{-\infty}^0 dz' M^4 e^{\kappa_b (z+z')}
e^{-|z-z'|q}
\end{equation}
\begin{equation}
I_4=\int_{-\infty}^0 dz \int_{-\infty}^0 dz' M^4 e^{\kappa_b (z+z')}
e^{(z+z')\/q}
\end{equation}
\begin{equation}
I_5=\int_0^\infty dz \int_{-\infty}^0 dz' M^2 N^2 e^{\kappa_b z'}
(z+z_0)^2 e^{-bz} e^{-(z-z')\/q}.
\end{equation}
These integrals are straightforward and the results are listed in
 Eqs.~(\ref{eq:I1})-(\ref{eq:I5}). The effective 2D interaction
 $U^{Q2D}(\vec R-\vec R')$ must reduce to ideal 2D case as
 $|\vec R-\vec R'| \rightarrow \infty$.
\begin{eqnarray}
\label{eq:app_lim}
\displaystyle\lim_{|\vec R-\vec R'|\to\infty}{U^{Q2D}
(\vec R-\vec R')}=\frac{e^2}{\epsilon_A |\vec R-\vec R'|} \left(1 +
 \frac{\epsilon_A -\epsilon_B}{\epsilon_A +\epsilon_B} \right)
 \left[ \int_0^\infty dz \,n(z) \right ]^2 + \frac{e^2}
{\epsilon_B |\vec R-\vec R'|} \nonumber\\
\left(1 + \frac{\epsilon_B -\epsilon_A}{\epsilon_A +\epsilon_B}
 \right) \left[ \int_{-\infty}^0 dz \, n(z) \right ]^2 +\frac{4 e^2}
{(\epsilon_A +\epsilon_B) |\vec R-\vec R'|} \int_0^\infty dz \,n(z)
 \int_{-\infty}^0 dz' \,n(z').
\end{eqnarray}
Using \begin{math} \int_0^\infty n(z) \, dz = 1-\int_{-\infty}^0
 n(z') \, dz' \end{math} in Eq.~(\ref{eq:app_lim}) leads to the
 desired result,
\begin{equation}
\lim_{|\vec R-\vec R'|\to\infty}{U^{Q2D}(\vec R-\vec R')}=
\frac{e^2}{|\vec R-\vec R'|}\frac{2}{\epsilon_A +\epsilon_B}.
\end{equation}
As a consequence the Fourier transform gives
\begin{equation}
\lim_{q\to 0}{U^{Q2D}(q)}=
\frac{2 \pi e^2}{q\, \bar \epsilon},
\end{equation}
so that \begin{math} \lim_{q\to 0}{F(q)}=1 \end{math} as can be
 observed in Fig.~1. This also indicates that all of the
 interactions are properly accounted in Eq.~(\ref{eq:app_U2}).

\pagebreak
\noindent
{\Large\bf List of Tables}
\begin{enumerate}
\item Fitting parameters $A$, $B$, $C$, and $D$ used in
 Eqs.~(\ref{eq:Gfit}) and (\ref{eq:Ffit}) as a function of $r_s$ for
 the characterization of the Q2D EL in a GaAs/AlGaAs
 heterostructure. The ionized acceptor density is $N_{depl}=
0.46\times 10^{11} \, \mbox{cm}^{-2}$. See Sec.~IV for the other
 parameters used for GaAs/AlGaAs system.
\item The constants used in Eqs.~(\ref{eq:Afit})-(\ref{eq:Dfit}) for
 different ionized acceptor densities, $N_{depl}$. The parameters
 characterizing the heterostructure are chosen suitable to the
 GaAs/AlGaAs system (see Sec.~IV).
\end{enumerate}
\pagebreak
\noindent
{\Large\bf List of Figures}
\begin{enumerate}
\item The Coulomb form factor $F$ and the  effect of the image terms
 as a function of wave number $q$ (in units of $k_F$) for the
 electronic densities $r_s$=0.8 and 20. The full lines apply to
 GaAs/AlGaAs heterostructure having $\epsilon_A$=13 and
 $\epsilon_B$=12.1. The dashed lines refer to the same system but
 with $\epsilon_A=\epsilon_B$=12.55 so that no image term appears.
 See Sec.~IV for the definition of $r_s$ and other parameters used
 for the system.
\item The local-field correction, $G^{Q2D}(q)$ of Q2D EL versus wave
 number $q$ (in units of $k_F$) for $r_s$ values 0.8, 1, 2, 3, 4, 5,
 10, 15, and 20. Solid lines: STLS and dashed lines: calculation
 using the fitted form for $G^{Q2D}(q)$; see Eq.~(\ref{eq:Gfit})
 with the values from Table~I.
\item The comparison of the full STLS Q2D local-field correction
 (solid lines) with that of Gold and Calmels' (dashed lines) given
 by Eq.~(\ref{eq:G_GC}) as a function of wave number $q$ (in units
 of $k_F$) for $r_s$=1 and 10.
\item Comparison of ideal 2D and Q2D inverse static dielectric
 function, $1/\epsilon(q,0)$ as a function of wave number $q$ (in
 units of $k_F$) for $r_s$=3. Solid lines: STLS, dashed lines: RPA.
 For Q2D EL, GaAs/AlGaAs heterostructure is used with $N_{depl}=
0.46\times 10^{11}\, \mbox{cm}^{-2}$.
\item The effect of the barrier height, $U_b$ on the inverse static
 dielectric function, $1/\epsilon(q,0)$ as a function of wave number
 $q$ (in units of $k_F$) for $r_s$=0.8. Solid line: $U_b=0.1\,
 \mbox{eV}$, dashed line: $U_b \to \infty$. Other parameters for the
 heterostructure are given in Sec.~IV.
\item The inverse static dielectric function of Q2D EL,
 $1/\epsilon(q,0)$ as a function of wave number $q$ (in units of
 $k_F$) for $r_s$ values 0.8, 1, 2, 3, 4, 5, 10, 15, and 20. Solid
 lines: STLS and dashed lines: calculation using the fitted forms
 for $G^{Q2D}(q)$ given by Eq.~(\ref{eq:Gfit}) and $F(q)$ given by
 Eq.~(\ref{eq:Ffit}) with the values in Table~I.
\item The electron distribution along the confinement direction in
 arbitrary units. The total electron density is $1\times 10^{12}\,
 \mbox{cm}^{-2}$. Other parameters are given in Sec.~IV. Solid line
 refers to the two subband populated calculation and the dashed
 line is based on the electrical quantum limit.
\item The normalized plasmon energy ($E_p/E_F\equiv 2\nu_p$) as a
 function of wave number $q$ (in units of $k_F$) for $r_s$ values 1,
 5, 10, and 20. Solid lines: STLS and dashed lines: calculation
 using the fitted forms for $G^{Q2D}(q)$ given by
 Eq.~(\ref{eq:Gfit}) and $F(q)$ given by Eq.~(\ref{eq:Ffit}) with
 the values in Table~I. The dotted line marks the onset of the
 electron-hole continuum.
\end{enumerate}

\vspace*{3cm}
\begin{table}
\begin{center}
\begin{tabular}{ c c c c c} 
\hline \hline    $r_s$ & $A$ & $B$ & $C$ & $D$\\
 \hline
 0.8	& 0.6243 & 0.4923 & 1.5462 & 1.2750 \\
 1.0	& 0.6549 & 0.5005 & 1.5079 & 1.1542 \\
 1.5	& 0.7250 & 0.5274 & 1.4342 & 0.9285 \\
 2.0	& 0.7857 & 0.5519 & 1.3950 & 0.7690 \\
 2.5	& 0.8380 & 0.5763 & 1.3644 & 0.6497 \\
 3.0	& 0.8794 & 0.5999 & 1.3512 & 0.5571 \\
 4.0	& 0.9405 & 0.6461 & 1.3274 & 0.4321 \\
 5.0	& 0.9779 & 0.6855 & 1.3264 & 0.3494 \\
 6.0	& 1.0012 & 0.7209 & 1.3356 & 0.2922 \\
 8.0	& 1.0225 & 0.7792 & 1.3683 & 0.2197 \\
 10	& 1.0294 & 0.8223 & 1.4097 & 0.1752 \\
 12	& 1.0305 & 0.8597 & 1.4545 & 0.1454 \\
 15	& 1.0295 & 0.9007 & 1.5014 & 0.1158 \\
 20	& 1.0257 & 0.9555 & 1.5185 & 0.0863 \\
 \hline \hline
\end{tabular}
\caption{}
\end{center}
\end{table}
\pagebreak
\vspace*{6cm}
\begin{table}
\begin{center}
\begin{tabular}{c c c c c c c c c c} 
\hline \hline    $N_{depl} \, (\mbox{cm}^{-2})$ & $a_1$ & $a_2$
 & $a_3$ & $b_1$ & $b_2$ & $b_3$ & $c_1$ & $d_1$ & $d_2$ \\
 \hline
 $0.146 \times 10^{11}$ & 0.6384 & 0.2213 & 0.5555 & 0.3023 & 4.9907
 & -0.0435 & 0.6891 & 4.3194 & 2.4659 \\
$0.46 \times 10^{11}$ & 0.6770 & 0.2794 & 0.6372 & 0.2575 & 2.6283
 & 0.1623 & 0.7914 & 2.7325 & 1.3674 \\
$1.47 \times 10^{11}$ & 0.6953 & 0.2302 & 0.6888 & 0.2253 & 1.4913
 & 0.3043 & 1.1675 & 1.6566 & 0.6898 \\
$4.69 \times 10^{11}$ & 0.6887 & 0.1802 & 0.7418 & 0.1978 & 0.6977
 & 0.4195 & 1.5437 & 1.0004 & 0.3518 \\
 \hline \hline
\end{tabular}
\caption{}
\end{center}
\end{table}

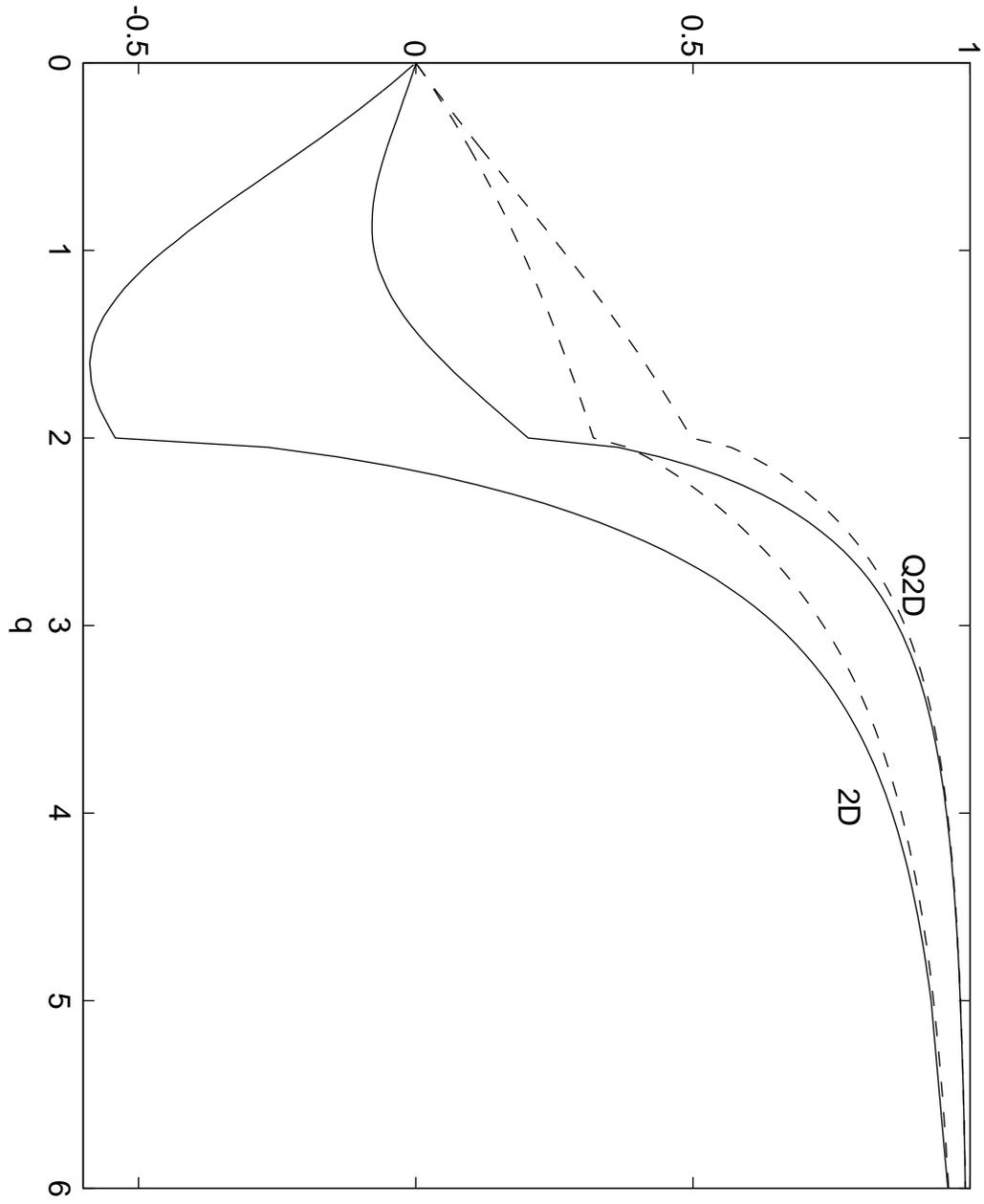
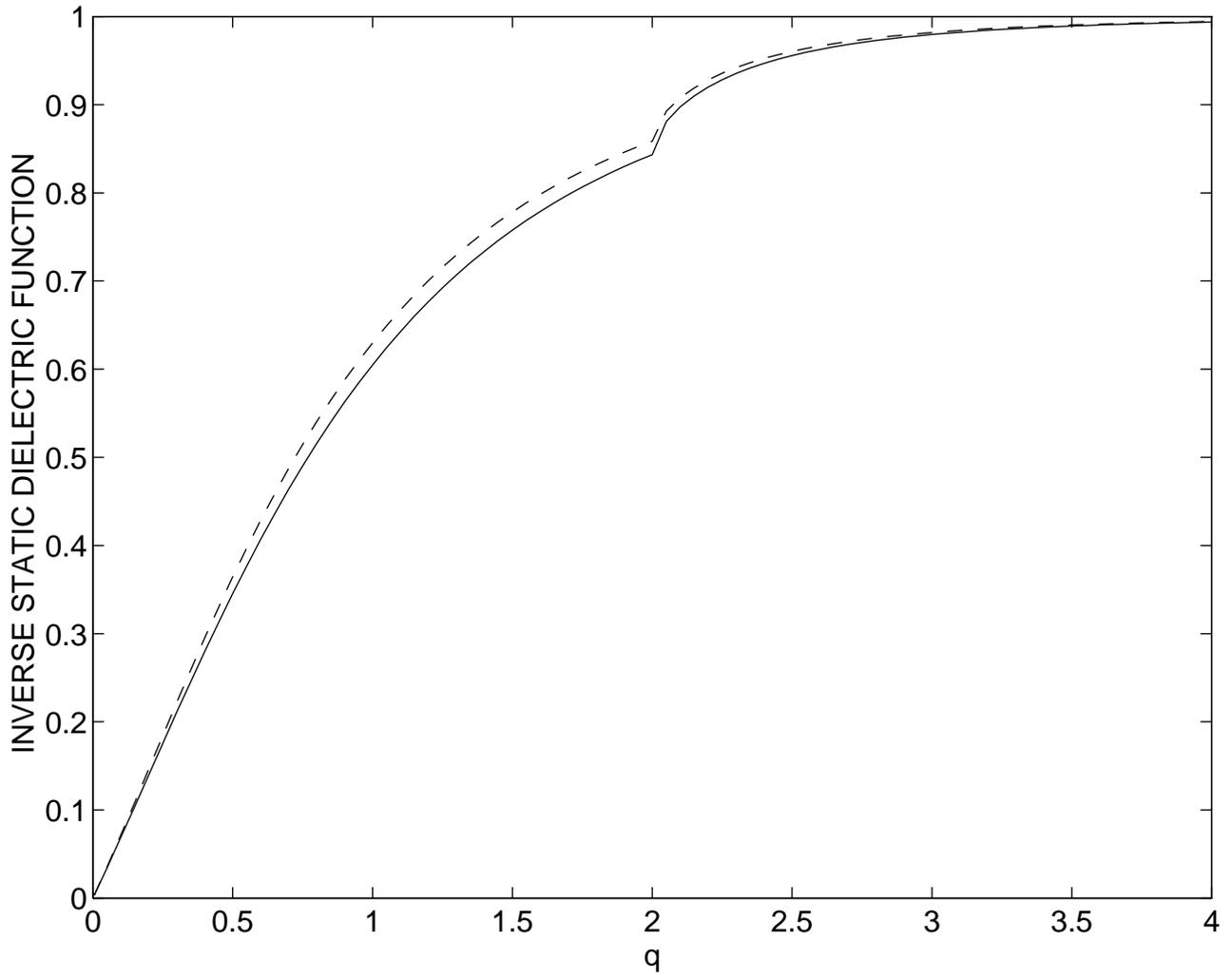
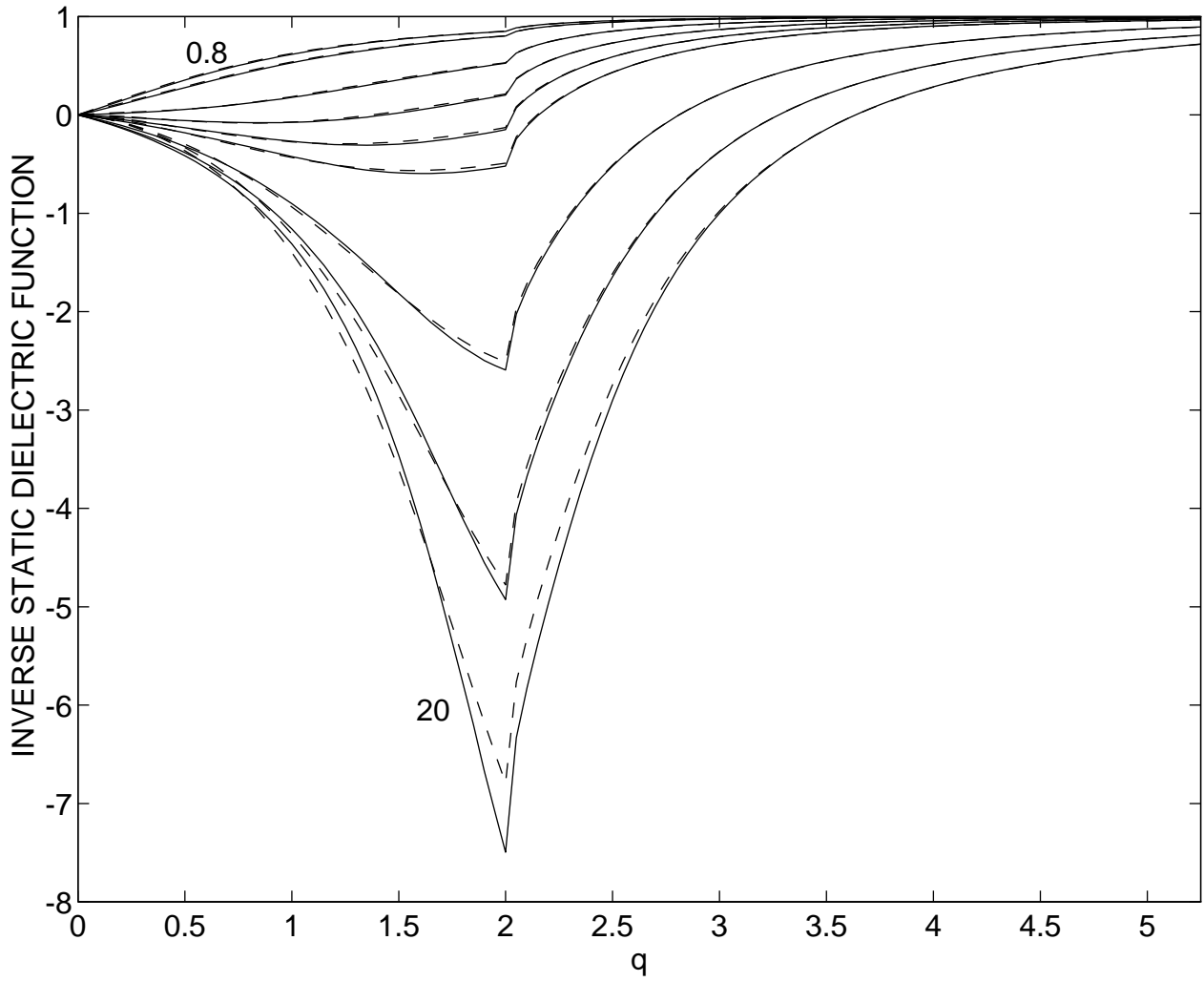
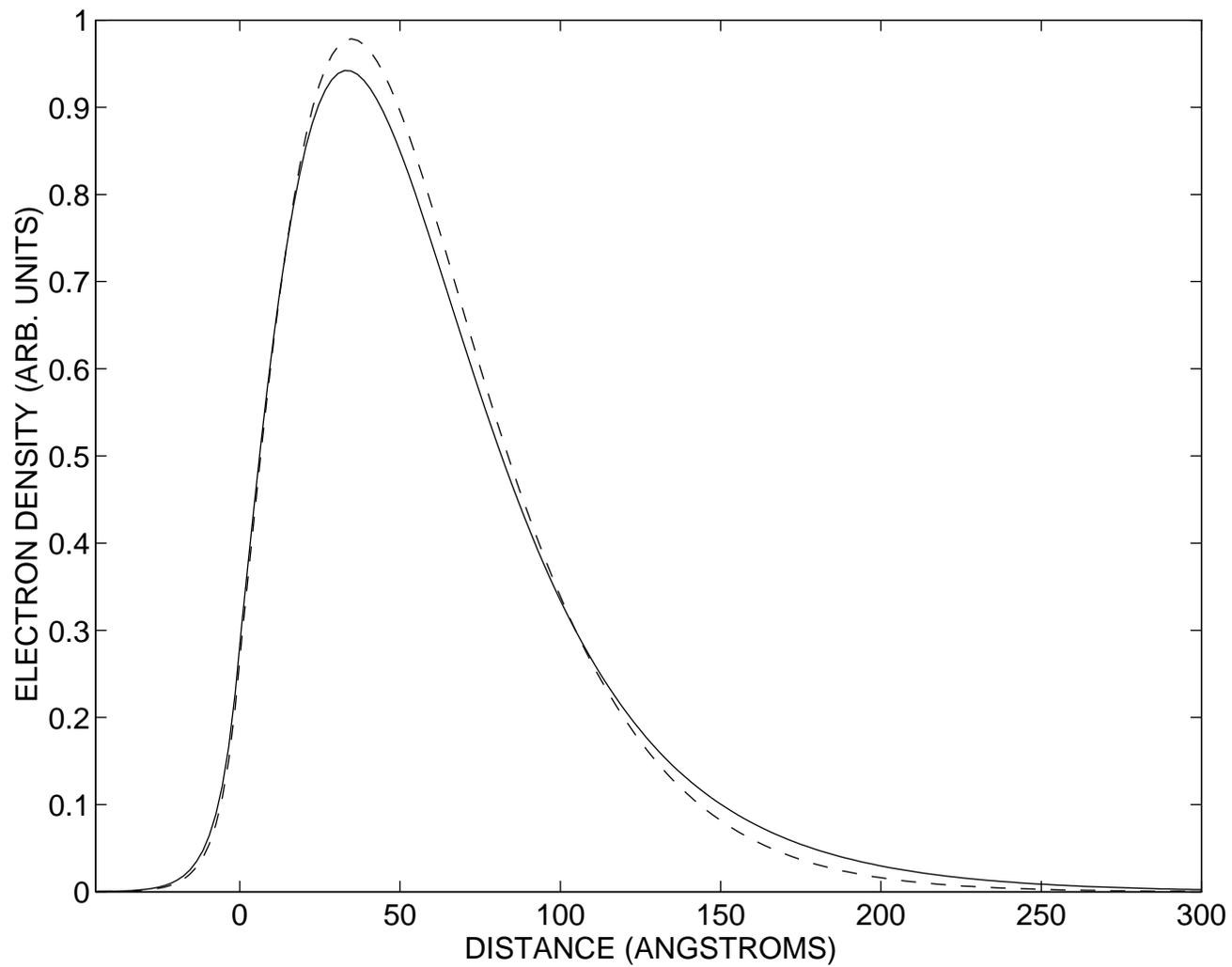
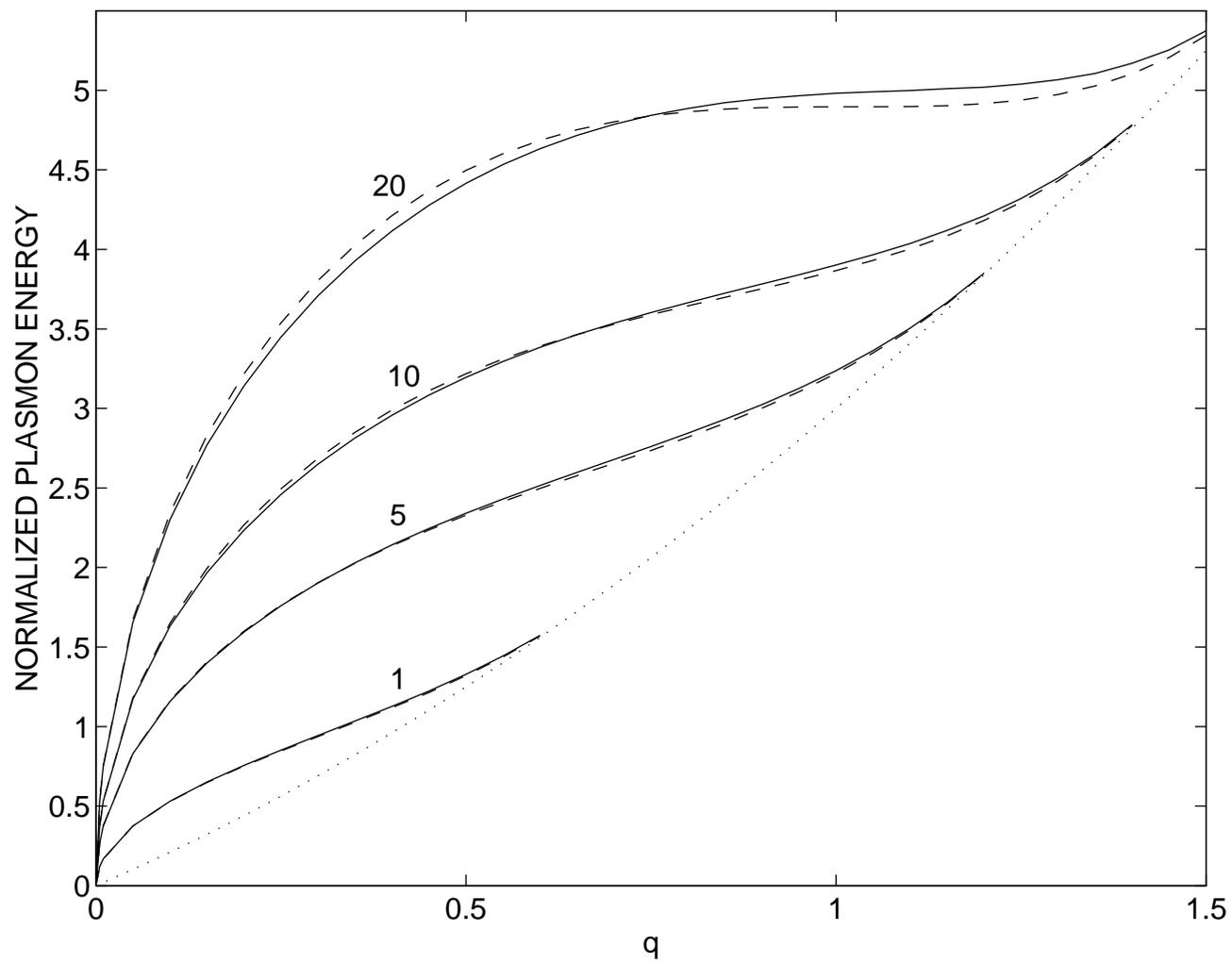
\end{document}